# Anisotropy of the in-plane angular magnetoresistance of electron-doed $Sr_{1-x}La_xCuO_2$ thin films

V. P. Jovanović, L. Fruchter, Z. Z. Li, and H. Raffy

*Laboratoire de Physique des Solides, UMR 8502-CNRS, Université Paris-Sud, 91405 Orsay, France*

**Abstract**

Signatures of antiferromagnetism (AF) in the underdoped $Ln_{2-x}Ce_xCuO_4$ ($Ln$ = Nd, Pr,…) family are observed even for doping levels for which superconductivity exists. We have looked for a similar property in a different electron-doped cuprate family, $Sr_{1-x}La_xCuO_2$, which consists of $CuO_2$ planes separated by Sr/La atoms, and is exempt of the possible influence of magnetic rare earth ions. We report in-plane magnetoresistance measurements in the normal state of underdoped, superconducting, *c*-axis oriented, epitaxial $Sr_{1-x}La_xCuO_2$ thin films. This probe is sensitive to spin arrangement and we find that the in-plane magnetoresistance, which is negative and does not saturate for $H \leq 6\,\text{T}$, exhibits an angular dependence when measured upon rotating a magnetic field within the $CuO_2$ planes. The analysis reveals a superposition of fourfold and twofold angular oscillations. Both of these increase in amplitude with increasing field and decreasing $T$ and appear below a temperature $T_{onset}$, which gets higher with decreasing doping levels. Our results demonstrate that these magnetoresistance oscillations, also observed for the $Ln_{2-x}Ce_xCuO_4$ ($Ln$ = Nd, Pr,…) family and attributed to an AF signature, are, without ambiguity, a property of $CuO_2$ planes. Besides, these oscillations vary with doping in an unusual way compared to previous results: fourfold oscillations are essentially present in the more underdoped samples while only twofold oscillations are visible in the less underdoped ones. This intriguing observation appears to be a consequence of spin dilution with increasing doping level.

PACS numbers: 74.72.Ek, 74.78.-w, 73.43.Qt

# I. INTRODUCTION

Undoped cuprates are antiferromagnetic (AF) insulators in which superconductivity gradually emerges upon doping the $CuO_2$ planes with holes or electrons. In the phase diagram two regions with distinct properties appear, one AF and the other superconducting (SC). The phase diagram is asymmetric with respect to the type of doping [1]: the AF region extends to a much higher doping level for electron-doped (*e*-doped) than for hole-doped (*h*-doped) cuprates while superconductivity is "stronger" for *h*-doped ones. This has been a subject of intensive studies involving neutron scattering or muon spin rotation ($\mu$SR) experiments, which are powerful tools for probing the spin subsystem, both for hole-doped [2] and electron-doped [3] cuprates. However some controversial results have been reported concerning the coexistence of magnetism and superconductivity in *e*-doped compounds [3, 4].

The phase diagram of *e*-doped cuprates has been established from studies carried on compounds belonging to a *single e-doped family*: $Ln_{2-x}Ce_xCuO_4$ (*Ln* = Nd, Pr,...). These compounds have complex crystallographic and spin structures and a study of compounds of another *e*-doped family, with a simpler structure, is of interest. The *e*-doped $Sr_{1-x}La_xCuO_2$ (SLCO) offers such a possibility. It only consists of a stack of square $CuO_2$ planes separated by (Sr,La) layers and can be considered as a model cuprate system [5]. The adjacent $CuO_2$ planes are not shifted by $(a/2, a/2)$ with respect to each other as for the *e*-doped family $Ln_{2-x}Ce_xCuO_4$ (*Ln* = Nd, Pr,...).and, more importantly, there is no magnetic ion in the structure. Since no single crystal of SLCO exists, the magnetism and spin arrangement in this compound have not been studied by neutron experiments. However $\mu$SR experiments carried on polycrystalline $Sr_{0.9}La_{0.1}CuO_2$ samples have shown that some magnetism survived in SC samples and that the magnetic volume fraction increased when the magnetic field increased [6]. Also recent tunneling measurements on optimally *e*-doped $Sr_{0.9}La_{0.1}CuO_2$ ceramic samples have revealed a hidden pseudogap inside vortex cores implying the existence of an order competing with superconductivity [7]. This competing order might as well be connected to antiferromagnetism, as the authors suggested.

Electronic in-plane magneto-transport measurements are well suited to probe the spin subsystem indirectly in cuprate thin films. An in-plane angular dependence of the magnetoresistance (MR) results from the correlation between the spin arrangements and the conduction electrons. Lightly *e*-doped *nonsuperconducting* cuprates $Ln_{2-x}Ce_xCuO_4$ (*Ln* = Nd, Pr,…) show significant coupling of spin and charge in various magnetotransport measurements [8–11]. In these systems symmetric fourfold oscillations of the angular MR (AMR) are observed when a constant magnetic field is rotated within conducting planes [8–11]. Lightly *e*-doped *superconducting* $Pr_{2-x}Ce_xCuO_4$ thin films also show fourfold oscillations of the AMR, possibly superimposed with twofold oscillations [12] while $La_{2-x}Ce_xCuO_4$ superconducting thin films only show twofold oscillations [13]. These oscillations persist up to optimal doping and have been attributed to intrinsic static antiferromagnetism coexisting with superconductivity [12]. However, the presence of a rare-earth magnetic atom (Nd or Pr) leads to some ambiguity when asserting that fourfold symmetry is uniquely due to $CuO_2$ planes.

With the motivation of probing the existence of some AF signature in the $CuO_2$ planes of SLCO, we have measured the in-plane normal state magnetoresistance ( $H \parallel ab$ ) of four *c*-axis oriented, epitaxial, underdoped superconducting $Sr_{1-x}La_xCuO_2$ thin films with different doping, and its angular dependence. We show that the in-plane MR under a magnetic field parallel to the $CuO_2$ planes is negative and exhibits, below some temperature increasing with decreasing doping, angular oscillations when rotating a field of given intensity within the planes. In the following we discuss these results which appear to confirm the presence of some magnetic order in the $CuO_2$ planes. The influence of doping, different from that reported for the other *e*-doped family, is discussed.

## II. EXPERIMENTAL

Single phase *c*-axis oriented epitaxial thin films of $Sr_{1-x}La_xCuO_2$ (SLCO) were deposited on heated $KTaO_3$ substrates by an rf magnetron sputtering technique and *in situ* reduced during the cooling stage of the preparation. A detailed description of the synthesis was given in Ref. 14. Sample 1 was prepared with $x = 0.10$. The three other ones (labeled 2, 3, and 4) were prepared with $x = 0.12$. Different doping states were also obtained by different *in situ* oxygen reduction conditions as explained in Ref. 14. These samples are underdoped: from sample 1 to 4, the critical temperature $T_c$ increases from 1 to 17 K while the resistance at 300 K decreases, indicating the increase in the number of carriers. X-ray diffraction (XRD) spectra [14] confirm that the films are epitaxial and single- phase. They are highly *c*-axis oriented as evident from the mosaicity of $0.1°$ or less, found from XRD $\omega$ scans [14].

The thin films were patterned, using electron beam lithography and chemical etching, in a standard six-contact resistivity bridge, the track being 0.35 mm wide and 0.80 mm long and parallel to the *a*-axis. Their thickness was from 60 to 120 nm. The transport measurements were performed with a four-probe low frequency ac method, in a magnetic field range: $0 \leq H \leq 6$ T and in the temperature range: $4.2 < T < 100$ K. The samples were mounted in such a way that the horizontal applied magnetic field was always parallel to the conducting plane and it was rotated around the film *c* axis (vertical). The current $I$ was typically 200 $\mu$A flowing along film *a* or *b* lattice axis and the voltage response was linear with the current. The temperature regulation was performed with a capacitive sensor, insensitive to magnetic field, and the temperature in zero magnetic field was measured with a Cernox thermometer. The measurements of the normal state MR, $\Delta\rho(H,T) = \rho(H,T) - \rho(0,T)$, were conducted with $H \parallel a \parallel I$ at given $T$. The angular dependence of the MR, $\Delta\rho(\theta,T,H)$, for given $T$ and $H$, where $\theta$ is the angle between the film *a* axis (or electric current, in most cases) and the magnetic field $H$ rotating within the film conducting plane, was also measured in the normal state.

## III. RESULTS

### A. In-plane magnetoresistance

Shown in Fig. 1(a) are the resistivity curves of the four samples studied, 1 to 4, with increasing $T_c$ ($1 \leq T_c(\rho=0) \leq 17$ K) all of them being underdoped as explained above. For all the samples, the resistivity, $\rho(T)$, displays a metallic behavior at high temperatures which transforms into an insulating one ($d\rho/dT < 0$) at low temperatures. This resistivity upturn can be related to disorder or possibly to spin scattering, as suggested by Dagan *et al.* [15] for *e*-doped $Pr_{2-x}Ce_xCuO_4$ thin films. The critical temperature $T_c^{mid}$ (taken at the transition midpoint) and temperature $T_p$ of the resistivity peak just above the transition of our samples increases monotonically with doping, quantified by the conductivity $\sigma_{300\,K}$ at room temperature [Fig. 1(b)]. In contrast, the temperature $T_{onset}$ below which oscillations in the AMR are seen for $H = 6$ T (described later in the text) decreases with increasing doping.

All samples have a negative in-plane MR, where $H \parallel a \parallel I$, at different fixed temperatures $T$ above the onset of superconductivity [Figs. 2(a)–2(c) for films 1-3]. In Figs. 2(d)–2(f) the corresponding values of the normalized MR at 6 T are plotted as a function of temperature. We notice here the decrease in the magnitude of the MR as the doping increases from sample 1 to samples 2 and 3 and its absolute value decreases linearly with increasing temperature. At low $T$ the MR tends to become positive [see Fig. 2(f)] below the temperature of the onset of superconductivity around $T_p$, due to the suppression of SC fluctuations by the magnetic field.[16]. In perpendicular magnetic field, the MR is always positive [16].

Roughly, the in-plane MR is proportional to $H^n$, where $n$ is typically between 1.4 and 2 for sample 1, while this exponent is close to 1.4 for samples 2 and 3 and does not vary with temperature.

### B. Anisotropy of the angular in-plane magnetoresistance

The most important results of this paper concern the normalized angular in-plane MR results, $\Delta\rho(\theta,T,H)/\rho(0,T,H)$. In Figs. 3(a) and 3(c) are shown the normalized AMR in a 6 T magnetic field at different fixed temperatures of samples 2 and 3, respectively. The two other samples have similar behavior: the MR curves of sample 1 and 4 resemble the one of sample 2 and 3, respectively. We observed fourfold oscillations of the AMR of samples 1 and 2 [Figs. 3(a) and 3(b)], the amplitude of which decreases when the temperature is increased. For the two other less underdoped samples, only twofold oscillations were visible [Fig. 3(c) and 3(d)]. For samples 1 and 2 it appears that the fourfold and twofold angular oscillations are both present. The data in Fig. 3 can indeed be decomposed in two sinusoidal components with two different periods ($\pi$ and $\pi/2$) and phases,

$$\rho(\theta) = C + A_2 \sin(2(\theta - \theta_2)) + A_4 \sin(4(\theta - \theta_4)), \quad (1)$$

where $C$ is a constant (at given $T$ and $H$), $A_2$ and $A_4$ are the amplitudes of twofold and fourfold oscillations, respectively, and $\theta_2$ and $\theta_4$ are their corresponding phases. The analysis of these parameters is given further below in Sec. III D of this section, while the fit to expression (1) of the AMR of sample 2 at 22 K is shown in Fig. 6.

In both cases of angular oscillations (twofold and fourfold) of the MR, whenever is $\theta$ equal to 0, $\pi$, or $2\pi$ then $H \parallel a \parallel I$ and we have a minimum of the oscillations (or a maximum of the absolute value of the MR). For films 1 and 2 a maximum of these fourfold oscillations (or a minimum of the absolute value of the MR) is seen whenever the angle between the current and the magnetic field is $\pi/4$ (the magnetic field is along the $Cu-Cu$ direction in this case) and a shallow minimum for $H \perp a$ ($H \perp I$). For samples 3 and 4 the corresponding maximum of twofold oscillations (or a minimum of the absolute value of the MR, or a maximum of the resistivity) appears when $H \perp a$ ($H \perp I$). We found no hysteretic behavior upon rotating the magnetic field from 0 to $2\pi$ and back.

**C. Influence of the magnetic field on the anisotropy of the angular magnetoresistance**

In order to establish the field dependence of the oscillations, we performed measurements of the AMR at a constant temperature in different magnetic fields for samples 1 and 3 [Fig. 4 (a) and 4(c)]. Their corresponding normalized amplitudes, $A_2/C$ and $A_4/C$, obtained from the fit to expression (1) are displayed in Figs. 4(b) and (d) as a function of the magnetic field. Both amplitudes increase monotonically when the magnetic field increases. A power-law fit [solid and dashed lines in Figs. 4(b) and 4(d)] reveals that the twofold and the fourfold amplitudes of sample 1 have almost a quadratic $H^2$ and a quartic $H^4$ variation with the field ($A_2 \propto H^{2.08}$ and $A_4 \propto H^{3.7}$), respectively. For sample 3 these exponents are significantly smaller, being 1.44 and 1.82 for twofold and fourfold amplitudes, respectively. This indicates that the fourfold oscillation should be more pronounced and easily visible at higher fields.

It is also observed that the amplitude of the AMR with essentially twofold oscillations [Fig. 4(c)] only depend on the component of the field $H\sin\theta$, perpendicular to the current direction or *a*-axis. As a matter of fact, all the data of the amplitude of the AMR, measured at different field intensities: 2, 3, 4, 5, and 6 T [see Fig. 4(c)], lie on a single curve when plotted as a function of $H\sin\theta$ [see inset of Fig. 4(d)].

### D. Analysis of the angular magnetoresistance

The fit of the data to expression (1) and its decomposition in two components are shown in Fig. 6. The temperature variation in these normalized amplitudes, $A_2/C$ and $A_4/C$ is given in Fig. 5. For both samples 3 and 4 [Fig. 5(c) and 5(d)] $A_4$ is much smaller than $A_2$ (if it exists at all). Close to the onset of the superconductivity, $A_2$ increases rapidly (data points out of scale), due to a small perpendicular component of the magnetic field, which is not perfectly parallel to the conducting planes. The magnitude of twofold oscillation of sample 4 [Fig. 5(d)] seems not to follow the apparent monotonic decrease with increasing $T_c$. This probably arises from the above mentioned field misalignment giving a perpendicular field component [16-18]. The amplitudes $A_2/C$ and $A_4/C$ decrease when the temperature increases and eventually disappear at certain temperature $T_{onset}$ (as defined in Fig. 5 and plotted in Fig. 1). We see that the temperature region where the oscillations exist is wider if the sample is more underdoped and $T_{onset}$ increases when doping decreases (see Fig. 1 for the variation in $T_{onset}$ with doping). Both phases do not change with temperature and their values are $\theta_2 \approx \pi/4$ and $\theta_4 \approx \pi/8$.

### E. Effect of the current direction

Another question regarding these oscillations is the role of the direction of the current or of the Lorentz force (for which we suppose that it is not related to superconductivity, i.e., vortices). To answer it, the direction of the current was changed from parallel to *a* axis to perpendicular to it (see insets of Fig.6). In these two different configurations (current along *a* and *b* axes), we had in both cases $H \parallel a$ for $\theta = 0$, but in the first case, $H \parallel I$ [current along *a*-axis, i.e. the main track, Fig. 6(a)] and in the second one $H \perp I$ [current along *b* axis, i.e., perpendicular to the main track, Fig. 6(b)]. A $\pi/2$ shift of the oscillations was observed (only visible on the twofold oscillations) when the direction of the current was changed to be parallel to the *b* axis of sample 2 (Fig. 6). Similar shift was observed for sample 3 (not shown). Now it becomes clear that in both cases the minimum of these curves (or the maximum of the absolute value of the MR) appears when $H \parallel I$, $I$ being parallel to equivalent Cu-O-Cu directions. This dependence on the current direction might be correlated with the cause of the twofold oscillations.

## IV. DISCUSSION

As first shown above, the in-plane MR is negative. This MR cannot be due to localization as it is only observed in parallel magnetic field nor to magnetic impurities. It appears that the resistance is reduced either with increasing field at given $T$, or decreasing $T$ under a given field. All these facts suggest that this negative MR is due to a reduction in spin scattering with increasing field and decreasing $T$. Also the effect is stronger for lower doping as the system is getting closer to the AF region: it appears below a temperature $T_{onset}$ [Fig. 1(b)] increasing with decreasing doping, which in our case is significantly lower than the resistivity upturn temperature (in Ref. 15, for *e*-doped $Pr_{2-x}Ce_xCuO_4$ SC thin films, it takes place at the same $T$ and the upturn was attributed to a spin scattering mechanism).

The in-plane normalized MR, $\Delta\rho(H,T)/\rho(0,T)$ (Fig. 2), shows no signature of a "spin-flop" transition (below which the MR is usually positive) : no saturation of the MR above some threshold magnetic field. This transition, seen for *e*-doped $Pr_{1.29}La_{0.7}Ce_xCuO_4$ (Ref. 8) and $Nd_{2-x}Ce_xCuO_4$ (Refs.10 and 11) [not seen in recent measurements on *e*-doped $La_{2-x}Ce_xCuO_4$ (Ref.13)] appears to be due to the presence of a magnetic ion in the structure and to a special, non-collinear [8, 19] AF spin arrangement in adjacent $CuO_2$ planes. In our infinite layer system, as indicated before, there is no $(a/2, a/2)$ shift between adjacent planes and the spin structure is unknown. Concerning *h*-doped cuprates, such as $YBa_2Cu_3O_6$ and $La_{2-x}Sr_xCuO_4$, the saturation of the MR above a threshold magnetic field was related to the establishment of the directional order of the stripes [20, 21]. As a consequence, an hysteresis appears in the MR of $YBa_2Cu_3O_6$ [20]. We did not observe any hysteretic behavior of the MR at a few fixed temperatures (around 20 – 30 K, in the normal state). For SLCO, there is no evidence for the existence of stripes, as far as we know.

The most important result is the anisotropy of the MR which seems to correlate with the crystalline structure of the material and reveal some anisotropic electronic or magnetic properties. The shape and the amplitude of AMR oscillations found in samples 1 and 2 in Fig. 3 are very similar to those of $Pr_{2-x}Ce_xCuO_4$ thin films in the $0.11 \leq x \leq 0.15$ doping range, although the authors did not give the decomposition of the AMR into two components [12]. Our results in SLCO confirm without ambiguity the fact that the MR oscillations are related to $CuO_2$ planes. These oscillations, larger when the doping decreases (closer to the AF region), can be ascribed to the presence of antiferromagnetism in the $CuO_2$ planes. There is a minimum of the scattering when the field is along the Cu-O-Cu direction, which may be related to the spin direction, which is not known in SLCO. The non-collinear arrangement known for $Pr_{2-x}Ce_xCuO_4$ in adjacent $CuO_2$ planes [19] might not be true for SLCO, as there is no magnetic ion, no $(a/2, a/2)$ shift between adjacent conducting planes and as the interplane spacing is smaller for SLCO (around 3.4 Å) than for $Pr_{2-x}Ce_xCuO_4$ (around 6.1 Å).

The magnitude (Fig. 5) of our fourfold oscillations is very small ($\sim 10^{-5}$ or $\sim 10^{-4}$) compared to those reported for undoped $Nd_{2-x}Ce_xCuO_4$ (Ref.11), (attributed to joint spin-flop and "spin valve" effect), but comparable to the one of nonsuperconducting $Pr_{2-x}Ce_xCuO_4$ (Ref. 9) attributed to the formation of stripe domains (without twofold oscillation contribution unlike in the present case). It is worth noting that pure fourfold oscillations in *e*-doped cuprates have only be reported in non superconducting samples.

Among non magnetic origins of the AMR, one might think that the fourfold oscillations somehow reflect the symmetry of a *d*-wave SC gap, which explains fourfold oscillations of the AMR in the mixed state of the *h*-doped $YBa_2Cu_3O_{7-\delta}$ [22]. Here, we are working in the normal state (small SC fluctuations) and even if there is such a possibility, we expect these oscillations to be also visible at high doping. Moreover, if due to SC fluctuations, the doping dependence would be homothetic to that of $T_c$ or $T_p$, which is not the case, as $T_{onset}$ follows an opposite trend. Also, the possibility that a pseudogap with *d*-wave symmetry may cause the MR oscillations observed in underdoped samples seems unlikely. Indeed there are some experimental evidence of the existence of a pseudogap in $Ln_{2-x}Ce_xCuO_4$ (*Ln* = Nd, Pr,…)). Tunneling experiments like in Ref. 23 have shown evidence of a low-energy normal-state gap, opening below a temperature close to $T_c$ much smaller than the temperature below which MR oscillations are seen. Besides, a high-energy pseudogap, shown in the optical conductivity of underdoped

Nd$_{2-x}$Ce$_x$CuO$_4$ single crystal, identified by the authors with the build up of AF correlations [Onose *et al.* 2004, Ref. 24], has an onset temperature $T^*$ very high, well above the Neel temperature $T_N$, and probably well above $T_{onset}$ in our SLCO samples.

The examination of different mechanisms lead us to conclude that the most plausible explanation of the fourfold oscillations of the in-plane AMR comes from the presence of an AF order in the CuO$_2$ planes. Concerning the origin of the twofold oscillations, it is less clear, as the rotational symmetry is broken. Recently, the in-plane AMR was measured on *e*-doped La$_{2-x}$Ce$_x$CuO$_4$ thin films (which does not contain magnetic atoms) (Ref. 13) and only twofold oscillations were found. Nevertheless the authors concluded that these twofold oscillations had also an antiferromagnetic origin.

Like in Ref. 13, we can argue that the twofold component does not come from an orthorhombic distortion, since the x-ray data show only (0 0 *l*) peaks and lattice parameters *a* and *b* are equal (within experimental error) [14]. The twofold component should not be of SC origin (whose influence decreases as doping decreases) [25].

It was shown in Sec. III E of the previous section that the resistance (or the scattering) is strongest when the magnetic field, parallel either to *a* axis or *b* axis, is perpendicular to *I* (Fig. 6). This scattering increase seems to depend only on the component of the field $H\sin\theta$ perpendicular to the current *I* [$\theta$ is the angle between the magnetic field and the current, see the inset in Fig. 4(d)]. Then, it is unexpected to observe, as we did, the decrease in the amplitude of twofold oscillations $A_2/C$ when *T* increases (and which eventually disappears at $T_{onset}$), if $A_2/C$ exists due to Lorentz force [26]. Moreover, in La$_{2-x}$Ce$_x$CuO$_4$ the twofold oscillations of AMR were found to be uncorrelated with the direction of the current (or to the Lorentz force) [13].

Finally, if the presence of an AF order is the cause of both fourfold and twofold AMR oscillations, the predominance of twofold oscillations and the disappearance of fourfold oscillations with increasing doping could be tentatively ascribed to a change in the in-plane spin order with spin dilution (electron doping in $e$-doped cuprates takes place in orbital $d$ of Cu, replacing $Cu^{2+}$ by $Cu^{+}$ spinless ion). One may imagine a scenario where one goes from a random repartition of spinless $Cu$ in a correlated two-dimensional AF structure at low doping, to a system with an ordered segregated phase, at higher doping, with a stripe-like one-dimensional (1D) AF structure (parallel to $a$ or $b$). The direction of the applied current would then select one direction or the other.

## V. CONCLUSION AND SUMMARY

By investigating the MR, in parallel magnetic field, of a series of lightly electron-doped SC epitaxial $Sr_{1-x}La_xCuO_2$ thin films, we have shown that their normal-state MR is negative, which is more likely a spin-dependent effect. The MR is anisotropic, which mirrors the crystalline structure and electronic and magnetic properties of the compound. The doping dependence of the in-plane AMR anisotropy is unique compared to the other $e$-doped cuprates: fourfold combined with twofold AMR oscillations were found in two the most underdoped films (1 and 2) and as the doping increases, only twofold AMR oscillations are essentially visible (films 3 and 4). The amplitudes of oscillations increase with increasing $H$ and decreasing $T$. The most probable origin of the fourfold oscillations is the presence of an AF order, as proposed for $Pr_{2-x}Ce_xCuO_4$ thin films [12]. According to our measurements, the magnetism appears to be really confined to the $CuO_2$ planes. The twofold component, always present in the oscillations, could also have an AF origin, as is proposed for $La_{2-x}Ce_xCuO_4$ thin films [13]. We tentatively suggest a scenario based on a segregation of spinless Cu with increasing doping leading to a stripelike 1D AF structure (parallel to *a* or *b* direction and selected by that of the applied current).

**ACKNOWLEDGEMENTS**

We thank F. Bouquet, Lj. Dobrosavljević-Grujić, M. Gabay, R. L. Greene, A. N. Lavrov, P. Monceau for helpful discussions. V. J. acknowledges support from the E. C. under an ESRT Marie Curie Program No. MEST-CT-2004-514307.

___________________________________________________________________

twofold amplitude $A_2/C$ increased by 80% at 14 K, close to SC transition, where the MR in perpendicular field increases rapidly when the field increases. The twofold amplitude $A_2/C$ was unchanged at 22 K. From $\Delta\rho(H)$ in a perpendicular field one can conclude that, already at temperatures $2-3$ K above $T_p$ the influence of a perpendicular field component on $A_2/C$ becomes insignificant.

negligible as one increases temperature 2 – 3 K away from $T_p$, i.e., the SC transition (Ref.16).

[26] The observed effect of the current direction does not necessarily mean that the Lorentz force is really related to the cause of oscillations of AMR. The Lorentz force is strongest when $H \perp I$ but in our case this is also the configuration where the current is parallel to one crystallographic axis ($I \parallel a$) and the magnetic field to the other ($H \parallel b$). In such a configuration one might imagine, for instance, a maximum (minimum) of the MR if the spins are parallel to crystallographic $a$ axis ($H \parallel b$) and the current flows along the same axis. Equally, a minimum (maximum) of the MR could be expected if the spins are parallel to crystallographic $b$ axis ($H \parallel a$) and the current flows along the $a$ axis.

**Figure captions**

FIG. 1 (Color online) (a) $\rho(T)$ curves of four different samples denoted with 1 to 4 by increasing critical temperature. (b) $T_c^{mid}$ (red circles), $T_p$ (black squares), and $T_{onset}$ (blue triangles with error bars, see also Fig. 5), as a function of doping ($\sigma_{300K}$). The shaded region is where the oscillations and negative MR are observed.

FIG. 2 (Color online) [(a)-(c)] The normalized MR as a function of a magnetic field parallel to the current direction, $H \parallel a \parallel I$, at constant temperatures for the three most resistive samples 1, 2, and 3, respectively. These temperatures are: (a) 12, 20, 28, 36, 44, 55, 65, and 75 K; (b) 19, 25, 30, 31, 35, and 41 K, and (c) 17, 18, 20, 24, 28, 32, 36, 43, 49, and 54 K. [(d)-(f)] The amplitude of the normalized MR as a function of temperature taken at 6 T from panels (a)-(c),

respectively. Errors were estimated from $\rho(T)$ with the assumption that the temperature stability was of few millikelvin and from possible contribution of perpendicular, out-of-plane component of magnetic field.

FIG. 3 (Color online) (a) Fourfold (film 2) and (c) twofold (film 3) oscillations of the isothermal normalized angular MR $\Delta\rho(\theta)/\rho(0)$ where $\theta$ is the angle between $a$ axis and 6 T magnetic field vector (which rotates in the conducting plane). Panels (b) and (d) are polar plots of the normalized AMR under 6 T at 16 K and 18 K of the two films shown in (a) and (c), respectively. Minimums of the oscillations correspond to the configuration in which $H \parallel a \parallel I$. The directions Cu-O-Cu ($a$ axis, $\theta = 0$ and $b$ axis, $\theta = \pi/2$) and $Cu-Cu$ ($\theta = \pi/4$) are indicated.

FIG. 4 (Color online) [(a) and (c)] The angular dependence of the normalized MR is plotted for increasing magnetic field values for samples 1 and 3, respectively. The magnetic field values are: (a) 1, 2, 3, 4, 4.5, 5, 5.5, and 6 T, and (c) 2, 3, 4, 5, and 6 T. [(b) and (d)] The normalized amplitudes of twofold (squares) and fourfold (circles) components as a function of the magnetic field and corresponding power fits $A/C = aH^n$ (solid and dashed lines, respectively). The inset in (d) shows that the data {shown in (c) in the form $[\rho(\theta) - \rho(0)]/\rho(0))$}, plotted as a function of $H\sin\theta$, lie on a single curve: $\Delta\rho = \rho(H\sin\theta) - \rho(0, H)$ (see the text for further information). The arrows indicate the maximum value of $\Delta\rho$ for the given magnetic field.

FIG. 5 (Color online) Normalized amplitudes of twofold (solid squares) and fourfold (open circles) oscillations, starting from the most resistive sample 1 in (a) to the least resistive sample 4 in (d). The onset of AMR oscillations is determined from the linear extrapolation to a temperature $T_{onset}$ [shown in Fig. 1(b)], where these amplitudes go to zero. Errors were estimated from the noise in $\Delta\rho(\theta)/\rho(0)$ (Fig. 2).

FIG. 6 (Color online) The normalized AMR $\Delta\rho(\theta)/\rho(0)$ in the configuration where the current is parallel to the (a) *a* axis and (b) *b* axis of the same film 2. The initial configuration, where $\theta = 0$, is given in insets. In both cases, the minimum of AMR oscillations under 6 T magnetic field at 22 K corresponds to a configuration in which $H \parallel I$. Squares represent the measured data while the solid curve is the fit. Dashed and dotted lines are the twofold and fourfold components of the fit, respectively.

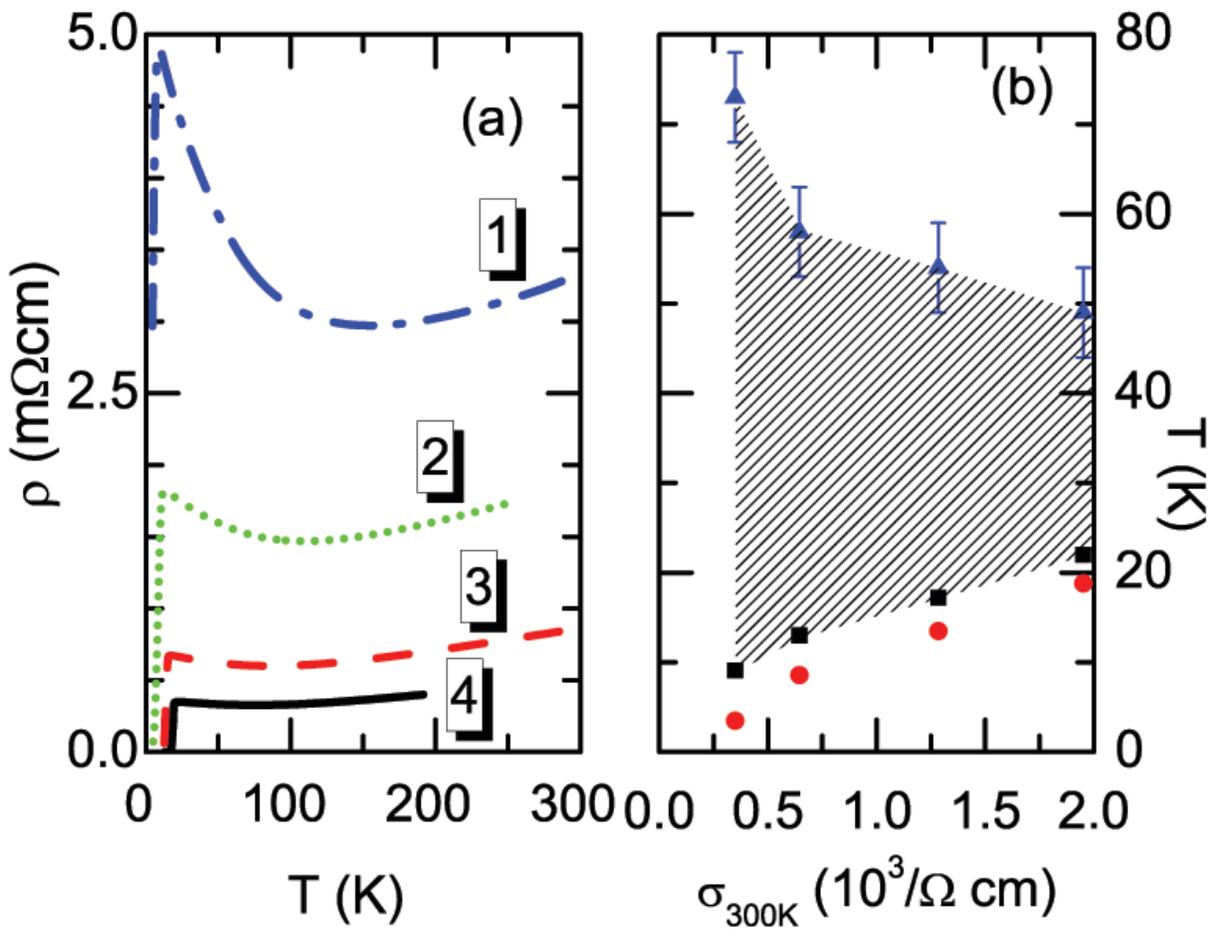

Fig 1. V. Jovanovic *et al*.

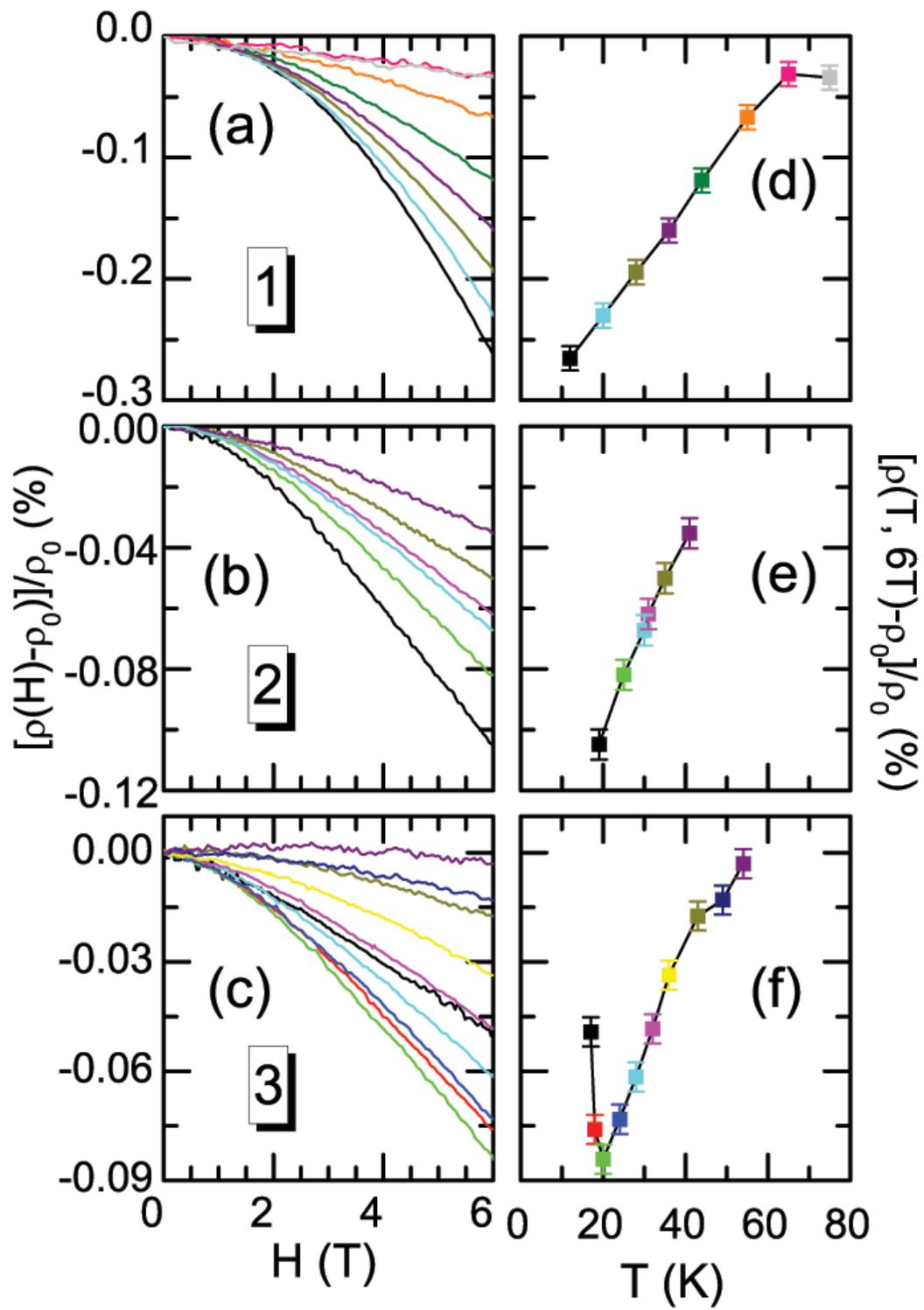

Fig 2. V. Jovanovic *et al*.

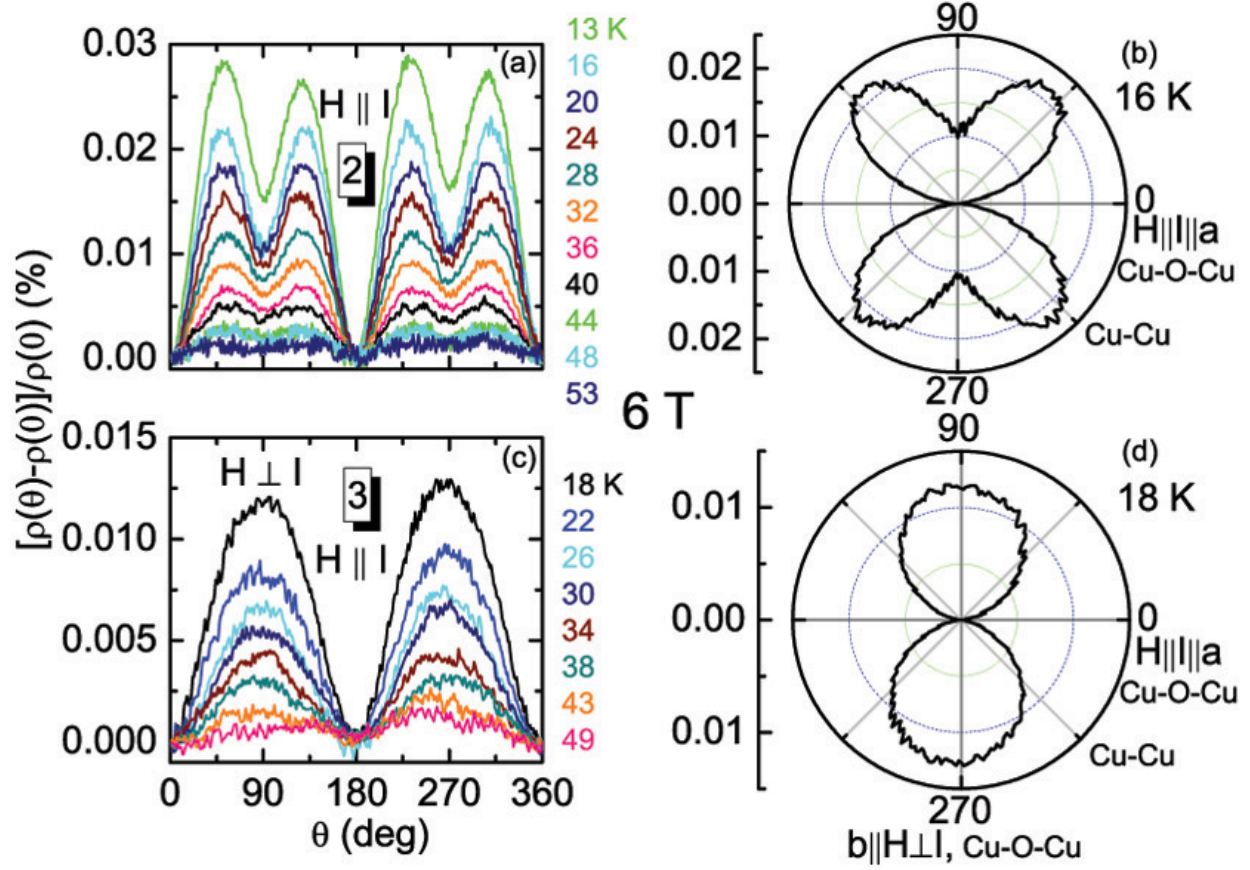

Fig 3. V. Jovanovic *et al*.

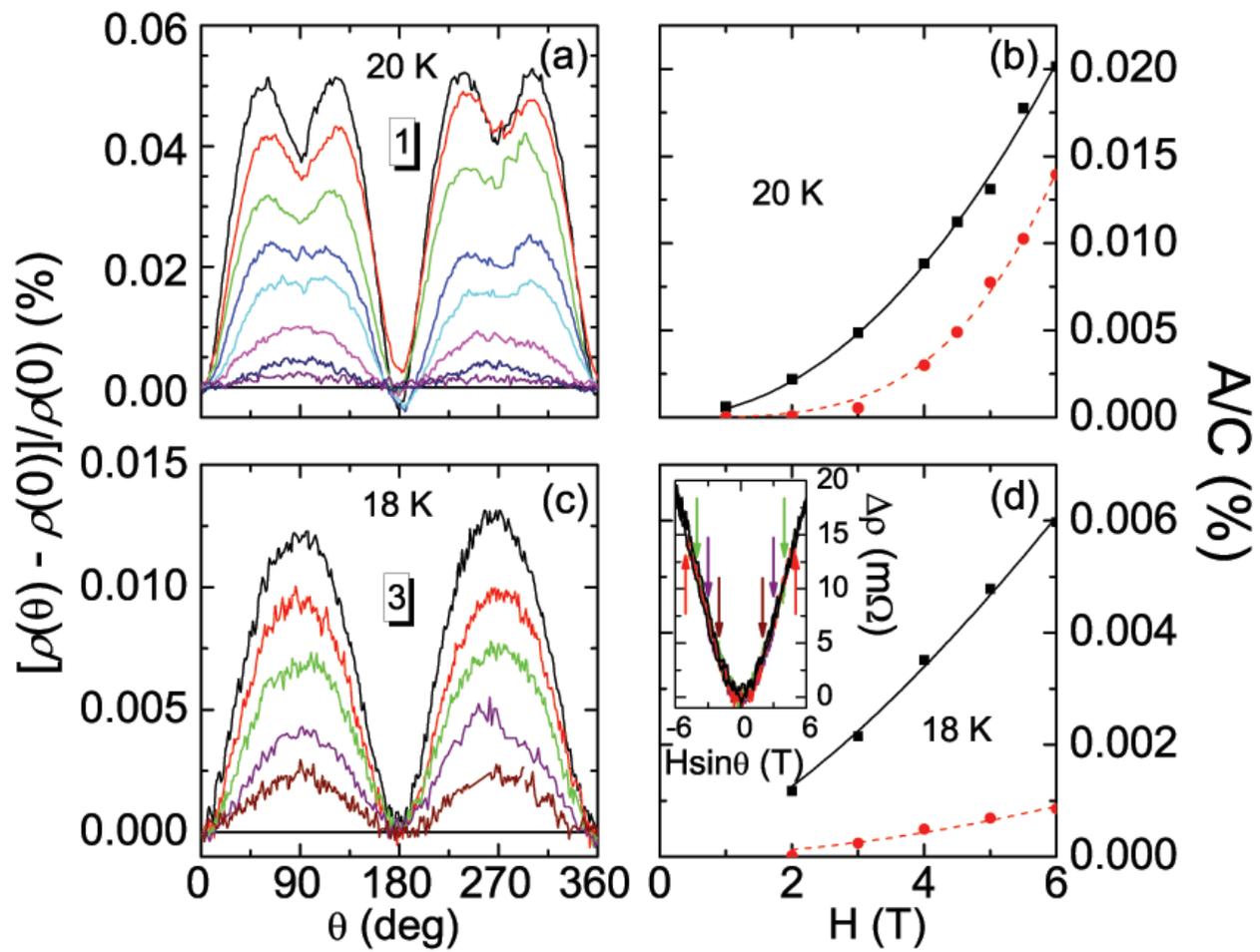

Fig 4. V. Jovanovic *et al*.

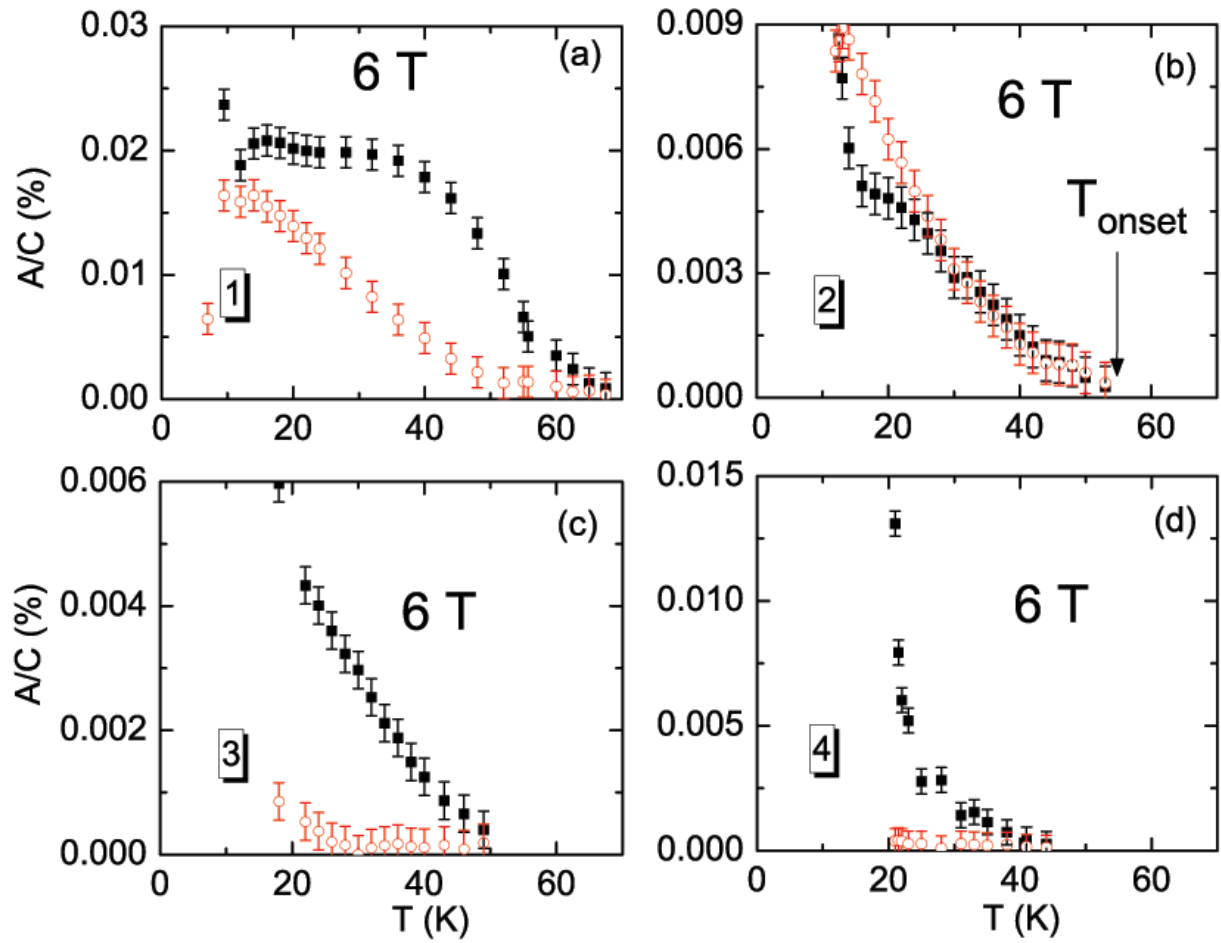

Fig 5. V. Jovanovic *et al*.

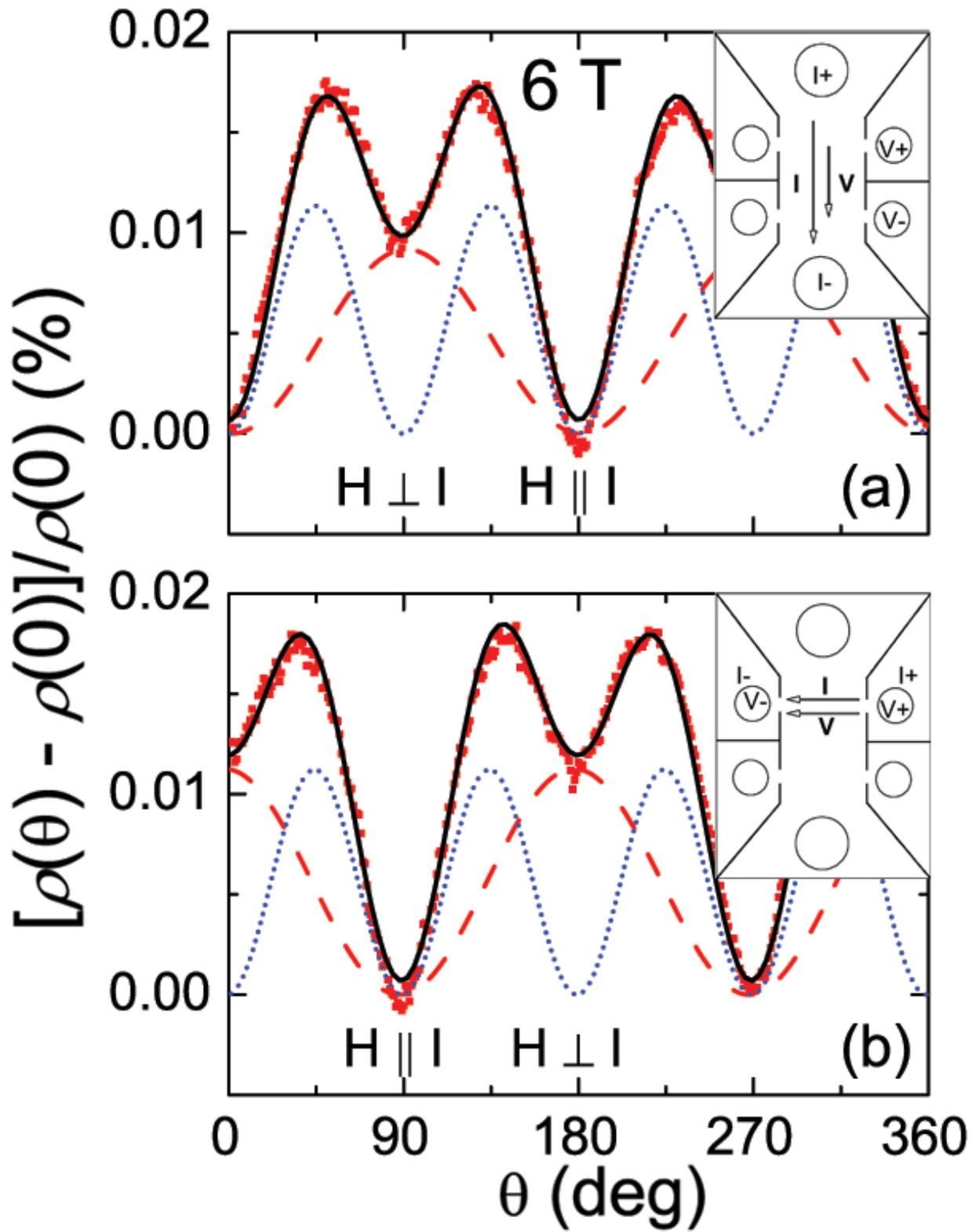

Fig 6. V. Jovanovic *et al*.